\documentstyle[12pt]{article}
\newcommand{\beq}{\begin{equation}}
\newcommand{\eeq}{\end{equation}}

\def\p1half{{\textstyle{{{p+1}\over{2}}}}}

\def\23phalf{{\textstyle{{{23-p}\over{2}}}}}

\begin{document}
\thispagestyle{empty}
\begin{titlepage}

\bigskip
\hskip 3.7in{\vbox{\baselineskip12pt
%\hbox{hep-th/0105244}
}}

\bigskip\bigskip\bigskip\bigskip
\centerline{\large\bf Deconfinement and the Hagedorn Transition in
String Theory}

\bigskip\bigskip
\bigskip\bigskip
\centerline{\bf Shyamoli Chaudhuri
\footnote{shyamoli@thphysed.org}
}
\centerline{214 North Allegheny St.}
\centerline{Bellefonte, PA 16823}
\date{\today}

\bigskip\bigskip
\begin{abstract}
Superseded and extended in hep-th/0105110 and hep-th/0208112.
\end{abstract}

\end{titlepage}

\section{Introduction}

\vskip 0.1in This paper has been superseded by hep-th/0105110 and
hep-th/0208112. The results in the published letter are not
incorrect, but the original presentation has been extensively
rewritten to clarify that thermal duality relations arise
naturally as a consequence of modular invariance in closed string
gases. Corresponding results for the type I open and closed string
gas are included in hep-th/0208112 giving a more unified picture
of the statistical mechanics of fermionic string gases, and of the
phase transition to the long string phase. The pedagogical case of
the free closed bosonic string gas appears in hep-th/0105110.
\end{document}